\documentclass[5p,times]{elsarticle}
\usepackage{lineno,hyperref}
\usepackage{mathptmx}  
\usepackage{amsmath}
\usepackage{amsfonts}
\usepackage{amssymb}
\usepackage{graphicx}
\usepackage{subfig}
\usepackage{soul}
\usepackage[dvipsnames]{xcolor}
\usepackage{romannum}
\usepackage[normalem]{ulem}
\usepackage{enumerate}
\usepackage{lscape}

\usepackage[normalem]{ulem}
\newcommand{\cmark}{\textcolor{green!30!black}{\ding{51}}}
\newcommand{\xmark}{\textcolor{red}{\ding{55}}}
\newcommand{\trot}[1]{\multicolumn{1}{l}{\rlap{\rotatebox{25}{#1}~}}}

\newcommand{\specialcell}[2][c]{%
	\begin{tabular}[#1]{@{}l@{}}#2\end{tabular}}

\usepackage{tikz}
\usepackage[english]{babel}
\usepackage{subfig}
\usepackage{paralist}
\usepackage{multirow}
\usepackage{booktabs}
\usepackage{caption}
\usepackage{framed}
\usepackage{listings}
\usepackage[ruled,linesnumbered]{algorithm2e}
\usepackage{xcolor, colortbl}    
\usepackage{soul}
\usepackage{pifont}
\usepackage{balance}
\hypersetup{
    colorlinks,
    linkcolor={black!50!black},
    citecolor={blue!50!black},
    urlcolor={blue!80!black}
}

\date{}

\def\BibTeX{{\rm B\kern-.05em{\sc i\kern-.025em b}\kern-.08em
    T\kern-.1667em\lower.7ex\hbox{E}\kern-.125emX}}
\begin{document}

\begin{frontmatter}

\title{EPICTWIN: An Electric Power Digital Twin for Cyber Security Testing, Research and Education}
\author{Nandha Kumar Kandasamy, Sarad Venugopalan, Tin Kit Wong, Leu Junming Nicholas}
\address{Singapore University of Technology and Design}

\begin{abstract}
Cyber-Physical Systems (CPS) rely on advanced communication and control technologies to efficiently manage devices and the flow of information in the system. However, a wide variety of potential security challenges has emerged due to the evolution of critical infrastructures (CI) from siloed sub-systems into connected and integrated networks. This is also the case for CI such as a smart grid. Smart grid security studies are carried out on  physical test-beds to provide its users a platform to train and test cyber attacks, in a safe and controlled environment. However, it has limitations w.r.t modifying physical configuration and difficulty to scale. 

To overcome these shortcomings, we built a digital power twin for a physical test-bed that is used for  cyber security studies on smart grids. On the developed twin,  the users can deploy real world attacks and countermeasures, to test and study its effectiveness. The difference from the physical test-bed is that its users may easily modify their power system components and configurations. Further, reproducing the twin for using and advancing the research is significantly cheaper. The developed twin has advanced features compared to any equivalent system in the literature. To illustrate a typical use case, we present a case study where a cyber attack is launched and discuss its implications.

\end{abstract}

\begin{keyword}
\texttt{Cyber Physical System, smartgrid, security, digital twin, test-bed.}
\end{keyword}

\end{frontmatter}


\section{Introduction}
\label{sec:intro}
It is impractical to test real world security vulnerabilities, countermeasures and performance impact on a live system without interrupting critical functions that may affect grid operations~\cite{khan2020}. Test-beds that can represent the realistically operation of critical infrastructures are of significant value to researchers and system operators. Description on many such test-beds are available in literature along with examples of their use cases for research in the design of secure smart-grids. From the survey in Cintuglu et al.~\cite{testbedsurvey}, it was observed that having the defense mechanisms evaluated in a physical test-bed facilitates smoother translation of developed technologies. However, the feasibility is relatively low for reproducing, scaling and modifying such test-beds. Particularly, for researchers working on cyber security technologies lower than TRL6 (technology readiness level 6), it is unrealistically expensive to reproduce such test-beds. Majority of the test-beds surveyed in a European commission report~\cite{Andreado2016}, were seen to typically cost around 2 million euros in initial capital expenditure. The overall costs including the operation and maintenance cost, restricts the use case to a smaller community of researchers.

In education, skills pertaining to the information security of a smartgrid (besides traditional power engineering curricula) are not commonly taught or a core focus of the course~\cite{Namboodiri2013,Strasser2014}. The  capex and opex for setting up a physical test-bed and maintaining it may be infeasible for smaller universities and educational institutions. Hence, there is a practical need for Digital Twins (DTs) that can offer sufficient fidelity of the physical system and emulated virtual network which is equivalent to a physical test-bed. Such DTs will eliminate the need for expensive test-beds for non-essential cases and enables the infrastructure to be available to a wider group of students, researchers and operator trainees. The feasibility to reproduce, scale and modify is another critical advantage of such DTs. 
\textit{Though we present a DT for a particular test-bed, the procedure used for creating the DT in this paper can be extended to any type of system whether a test-bed or an actual plant.}

The DT developed and presented in this paper is a digital equivalent of the physical micro-grid test-bed used in Adepu et al~\cite{Sridhar2019} and is called as Electric Power and Intelligent Control (EPIC) test-bed. The EPIC test-bed allows its users to validate the security and safe operation of critical components. We emphasize that our twin is primarily focused on security training and research. Its behavior is closely modeled to approximate the physical test-bed --- to study its impact on security. It provides users a more realistic view and reduce the gap between physical and simulated test-bed environments for security testing.
The digital components (corresponding to a physical system) and its internal configuration  are easy to modify, scale and interconnect. The user is not expected to have in-depth programming knowledge. A system expert will be able to duplicate existing modules, change parameters, and plug-and-play into the DT.
Our DT supports a number of communication protocols (e.g., mms, goose, mqtt, opc-ua, modbus). The network topologies, traffic flow and other parameters can be changed easily. It also allows external software modules to be added to our DT.
\textit{A modular and flexible approach is incorporated into our DT to support $\lq$end to end security testing$\rq$ for different test cases}.

Our DT allows users to  deploy attacks and countermeasures. Some of the capabilities (but not limited to) of our DT are listed below.

\begin{figure*}[!htb]
\centering
\includegraphics[width=0.9\linewidth]{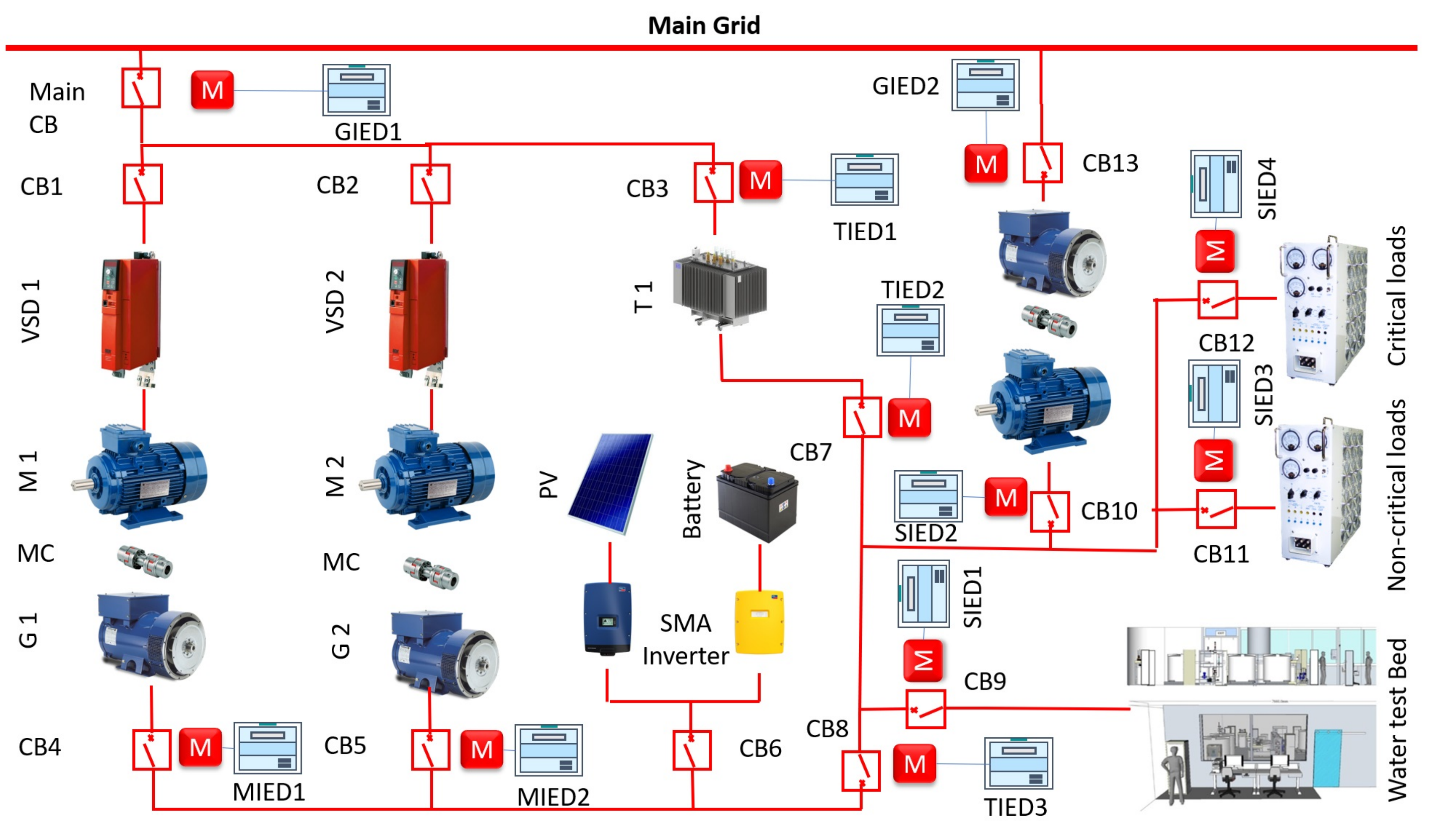}
\caption{Electrical layout of the EPIC (actual) test-bed. Electrical power lines are shown in red color lines. \small{MC - Mechanical coupling, CB - Circuit Breaker, IEDs - intelligent electronic devices, the prefix G, M T and S stands for Generation, Micro-grid, Transmission and Smart Home respectively}.}
\label{fig:EPIC_electrical_layout}
\end{figure*}

\begin{figure*}[!htb]
    \centering
    \includegraphics[scale=0.65]{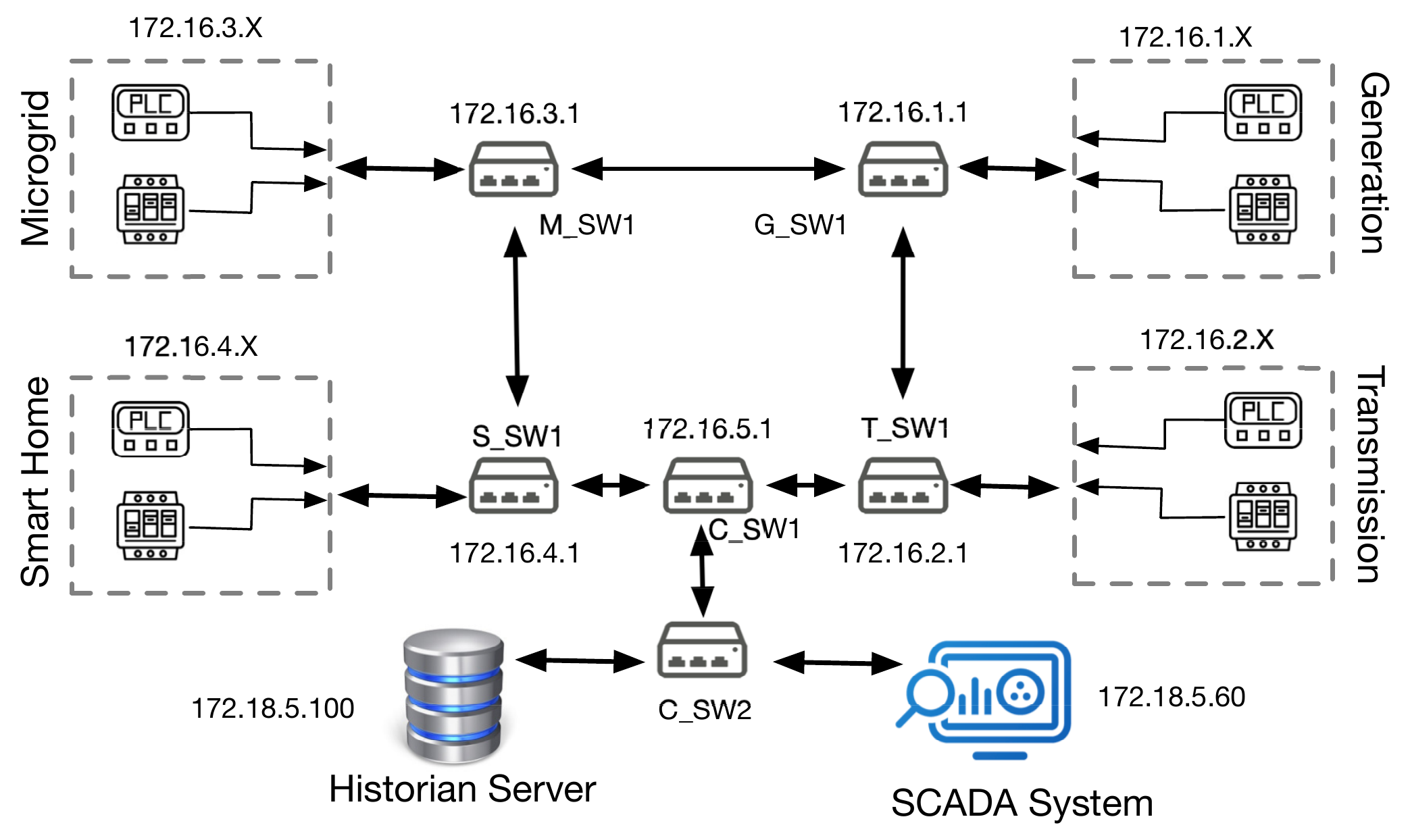}
    \caption{A simplified network diagram for EPIC test-bed is shown. This is to help the associated network traces. IP addresses in the network traces correspond to respected devices shown in the figure. Each dotted box represents a subnet with the respective IP addresses. A X in the IP address means that a device in that subnet would have the similar subnet mask and then unique X as its own IP. Typical devices are PLCs and Intelligent Electronic Devices. The connection $M\_SW1-->S\_SW1-->C\_SW1-->T\_SW1-->G\_SW1-->M\_SW1$ is a High-availability Seamless Redundancy (HSR) ring.}
    \label{EPIC_network_fig}
\end{figure*}

\begin{figure*}[!htb]
    \centering
    \includegraphics[width=\linewidth]{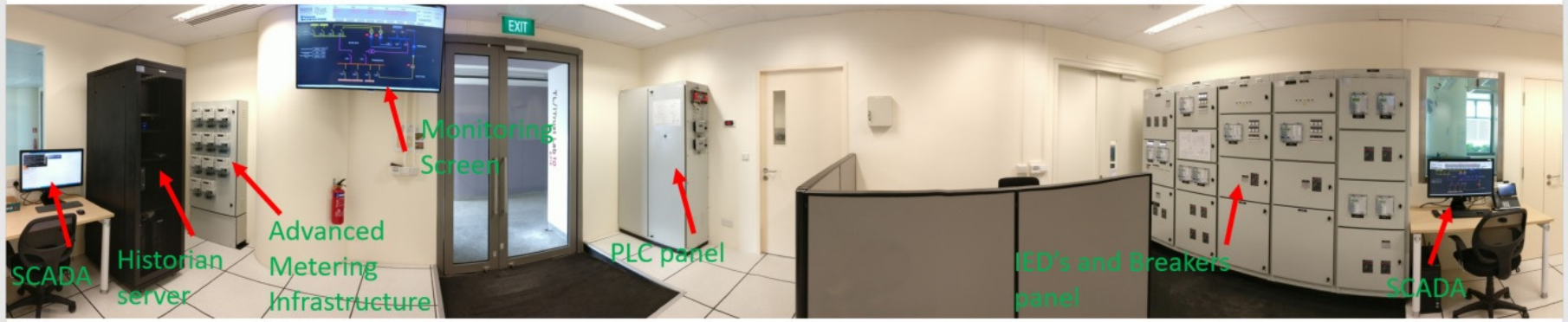}
    \caption{EPIC control room, 360-degree view.\small{The picture shows different physical components including the SCADA workstation, IED and Breaker panel, PLC panel, Historian server, Advanced Metering Infrastructure (AMI)  and the Monitoring screen.}}
    \label{control_room_fig}
\end{figure*}

\begin{figure*}[!htb]
    \centering
    \includegraphics[scale=0.4]{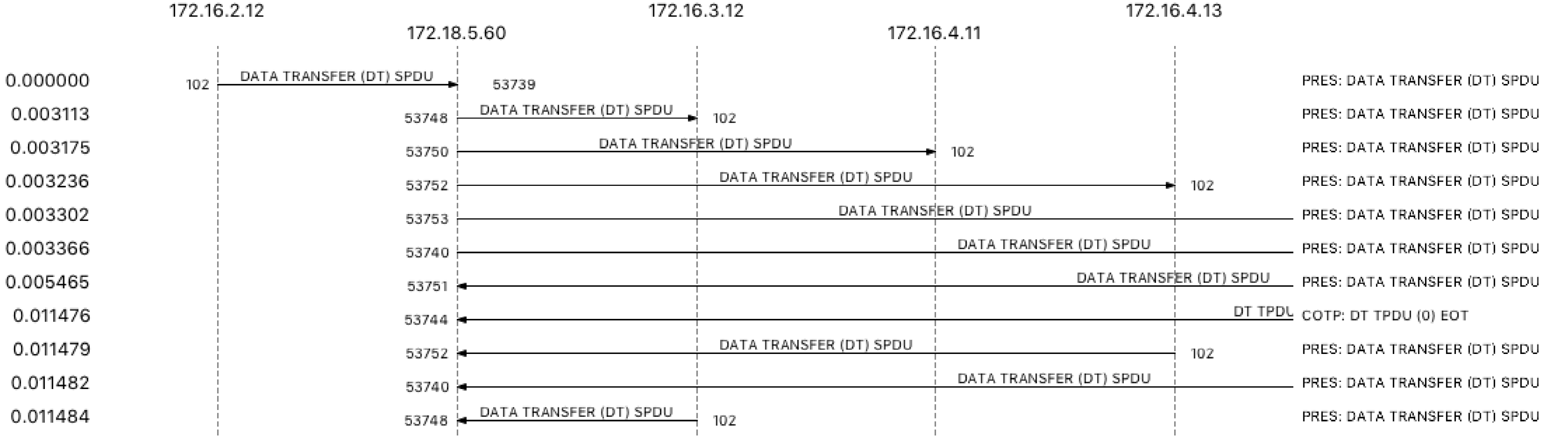}
    \caption{A sample network traffic flow between different devices.}
    \label{network_sequence_fig}
\end{figure*}

\begin{compactenum}[i)]
    \item No hardware components are required.
    \item At the service layer, the software (application) under consideration may be tested for vulnerabilities.
    \item It is able to send/receive data via a (virtual) network, enabling analysis on network traffic with physical process data. 
    \item  At the network layer, we use virtual switching with a switching stack for hardware virtualization. It supports standard network protocols allowing its user to easily test out and validate attacks.
    \item At the device layer, a firmware emulator may be used to test and patch device exploits or alternately connect to  physical components in a hybrid setup.
    \item Since separate data bus is used for the physical signal, integrating with physical test-beds for creating hybrid-twins is also enabled.
\end{compactenum}
Our contributions are unique from the following aspects,
\begin{compactenum}
    \item The test-beds available in literature are either hardware test-beds or digital test-beds and in some cases a hybrid. There are no known systems available in literature which are a digital twin of an existing cyber-security test-bed or plant. Hence, the system presented in the paper, to the best of our knowledge is the first of its kind.
    \item The paper presents a digital equivalent for an existing physical test-bed with a procedure that can be used for twining any test-bed/actual plant meant for cyber-security studies. 
    \item We  provide an attacker designer (AD) and attack launcher (AL) for launching systematic attacks on the twin. The AD and AL are unique from the aspect that the researchers without a background in offensive security testing could utilize them to validate and improvise the developed defense mechanisms. E.g., security researchers with expertise in machine learning working on grid anomaly detection.
\end{compactenum}

The rest of the paper is organized as follows, the background required for understanding paper is presented in Section~\ref{sec:background} followed by the architecture of the twin in Section~\ref{sec:DTarch}. A case study is presented in Section~\ref{sec:casestudy} and the discussion on the features is presented in Section~\ref{sec:discussion}. Section~\ref{sec:related} presents the related work and the conclusions are drawn in Section~\ref{sec:conclusions}.

\section{Background}
\label{sec:background}
\subsection{Use Cases}
\label{sssec:Usecases}
In our DT, the following use cases can be realized and these use cases are listed as critical requirements for DTs by Bergman et al.\cite{Bergman2009} \& Ashok et al.\cite{Ashok2016}:

\begin{compactenum}[i)]
\item Training operator trainees, students and researchers for operating and using a test-bed or actual plant,
\item Model \& library development for a test-bed or actual plant,
\item Threat \& vulnerability assessment of different components in a test-bed or actual plant,
\item Threat analysis \& mitigation for different components and vulnerabilities, and
\item Development, testing, deployment and rapid prototyping of different technologies, control strategies and topologies.
\end{compactenum}

\subsection{Architecture of Physical Test-Bed Considered for Twining}
\label{sssec:testbedarch}	
\subsubsection{Test-bed overview} 

The EPIC test-bed~\cite{Sridhar2019} mainly consists of four zones namely, generation, micro-grid, transmission and smart home.
All the four zones are equipped with intelligent electronic devices (IEDs) to collect Current, Voltage, Power and Frequency for the three phases buses. Further, advanced metering infrastructure (AMI) meters are also physically co-located with IEDs and collect Current, Voltage, Power and Frequency for all the three phases at different electrical nodes. 
\begin{itemize}
    \item \textbf{Generation:} Generation stage is driven by electric motors connected to the main power supply. 
    
    \item \textbf{Micro-Grid:} Photo-voltaic (PV) cells, inverters and batteries compose this stage to supplement the generation of power.   
    
    \item \textbf{Transmission:} This stage is composed of buses to transport power to the smart home unit.
    
    \item \textbf{Smart Home:} Programmable load banks containing RLC\footnote{Resistor-Inductor-Capacitor} loads represents a home load environment. Besides, there are two water test-beds~\cite{Mathur2016,Mujeeb2017} connected to the EPIC test-bed  as the load.
    
\end{itemize}

\subsubsection{Electrical Layout}
The test-bed consists of the following components as shown in Figure.\,\ref{fig:EPIC_electrical_layout}. It has
\romannum{1}) two conventional generators (10KVA each) driven by 15kW VSD driven motors to represent the conventional combination of prime-mover and generator. 
\romannum{2}) a 34kW solar PV system is available along with an 18kW battery system to represent power generation from the intermittent Renewable Energy Source (RES).
\romannum{3}) a 105kVA 3 phase voltage regulator for representing power supply from a transmission system. 
\romannum{4}) two load banks capable of emulating 45kVA load to represent critical and non-critical loads. It can supply power to the other two water test-beds.
\romannum{5}) a 10kW motor-generator load to represent spinning load and 
\romannum{6}) industrial standard Molded Case Circuit Breakers used for short-circuit protection and switching functions.

A picture of the control room is shown in Figure.~\ref{control_room_fig}. A digital equivalent of this test-bed will be presented in the following sections.

\begin{figure*}[h]
    \centering
    \includegraphics[width=0.9\linewidth]{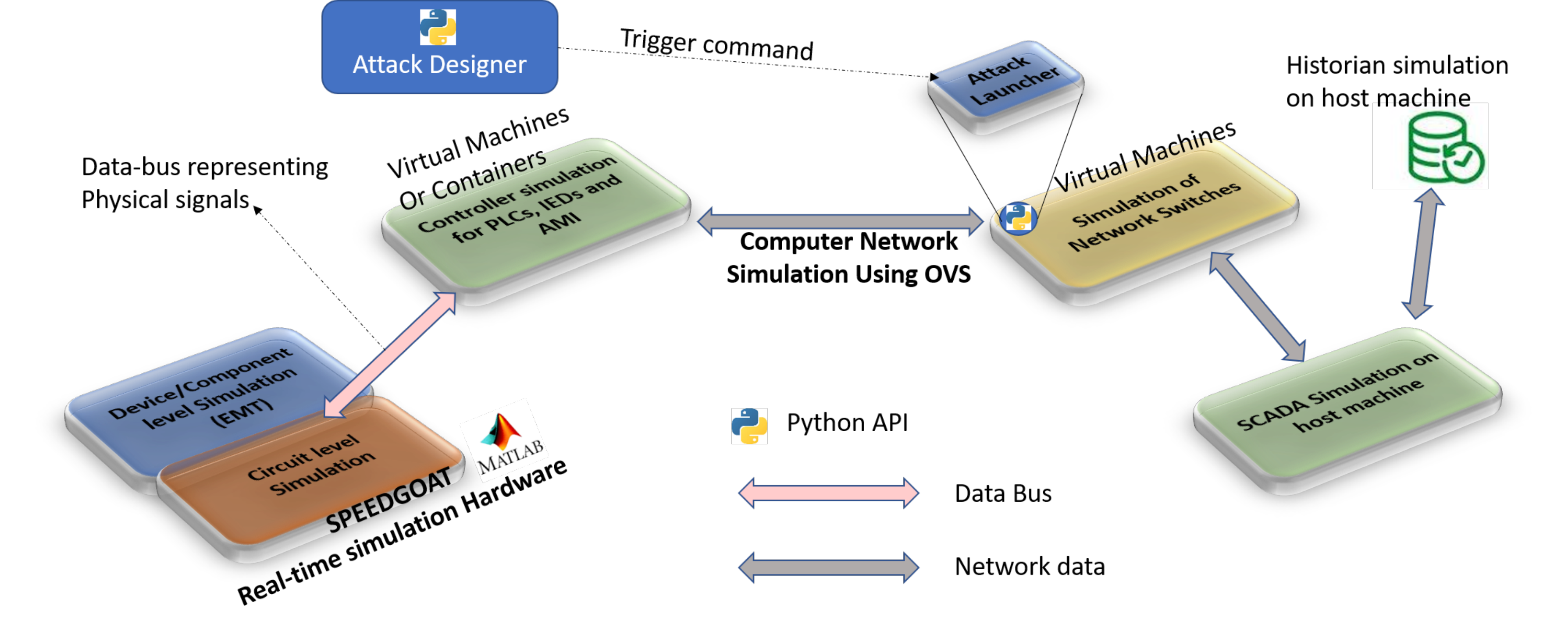}
    \caption{Architecture of the Digital Twin. \small{Figure.~\ref{fig:EPIC_electrical_layout} is simulated on SPEEDGOAT and Figure~\ref{EPIC_network_fig} is simulated using a workstation. Different software are used to achieve the functions of the whole system.}}
    \label{DT_arch_fig}
\end{figure*}

\begin{figure*}[!htb]
	\centering
	\includegraphics[width=0.9\linewidth]{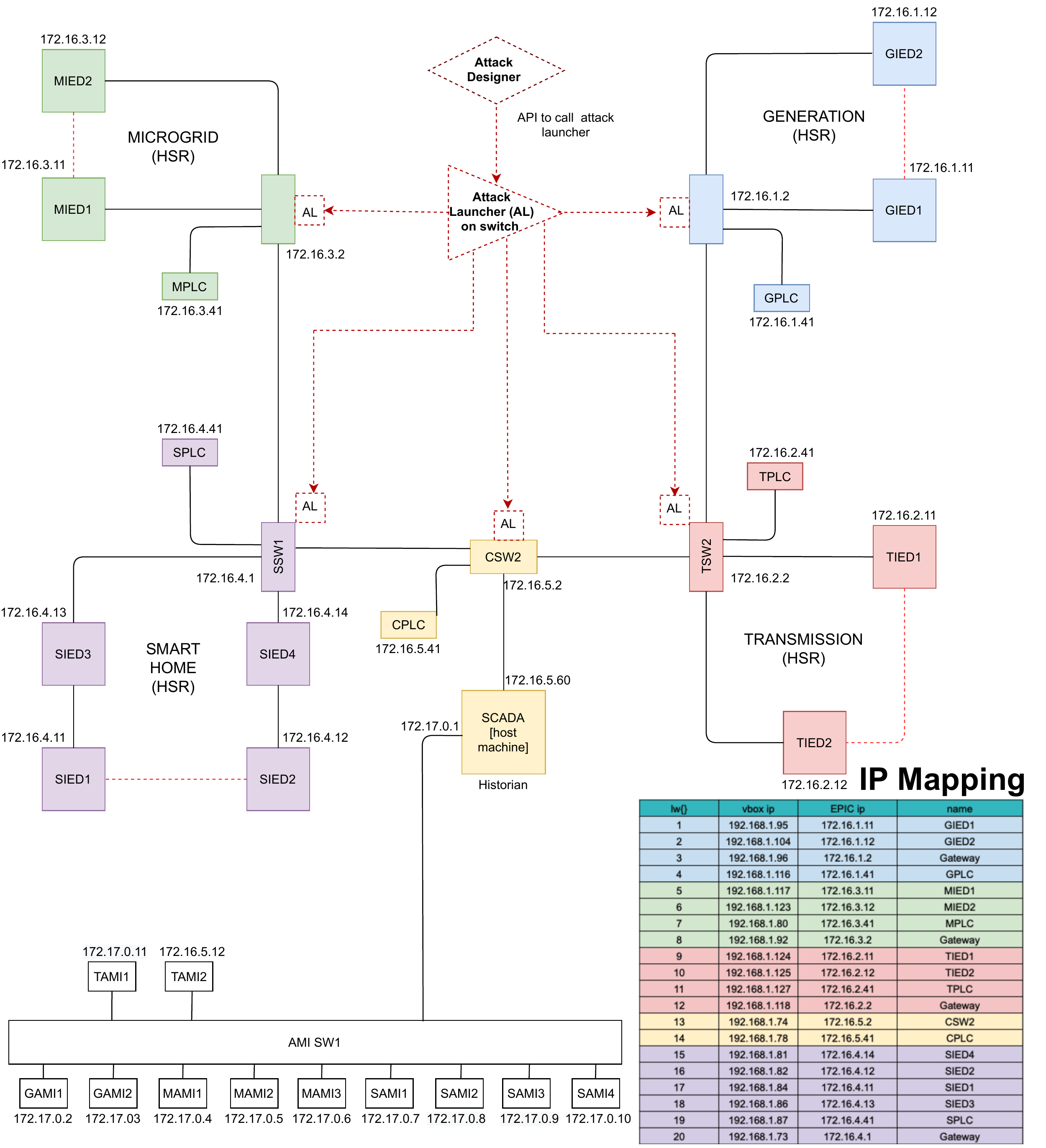}
	\caption{EPIC digital twin network diagram. The AMI are implemented on docker containers, The  IEDs and PLCs run on VMs. The connections are made via virtual switches. The SCADA runs on the host machine. The IP map gives the mapping between 192.168.1.x IPs for VMs communication with the external MQTT broker (not shown in Figure) and  172.16.x.x IPs  for communications within the switched network. Refer Figure.~\ref{fig:iedmodel} for detailed structure of the individual IEDs e.g., GIED1.}
	\label{fig:epicsploit}
\end{figure*}

\subsubsection{Communication Layout}
Figure~\ref{EPIC_network_fig} shows the communication network architecture in EPIC test-bed. It shows four major control zones, namely, power generation, transmission,  micro-grid,  and  smart home. All four zones have IEDs and other devices controlled by dedicated Programmable Logic Controllers (PLCs). For example smart home also contains AMI meters that can communicate with the PLCs and then can route the data through a central switch to the historian server and a SCADA workstation.  The map of IP addresses of the specific devices in the test-bed to the twin is shown in Figure.~\ref{fig:epicsploit}. For details on the test-bed related information, an interested reader is referred to the EPIC test-bed papers~\cite{Sridhar2019,datasetpaper}. Figure~\ref{network_sequence_fig} shows the sequence of communication between the different devices in the EPIC network. In the sequence diagram it can be seen that the SCADA\footnote{Supervisory control and data acquisition.} workstation (IP: 172.18.5.60) holds a central position and send data transfer requests to the rest of the PLCs.

\begin{figure*}
	\centering
	\includegraphics[width=0.8\linewidth]{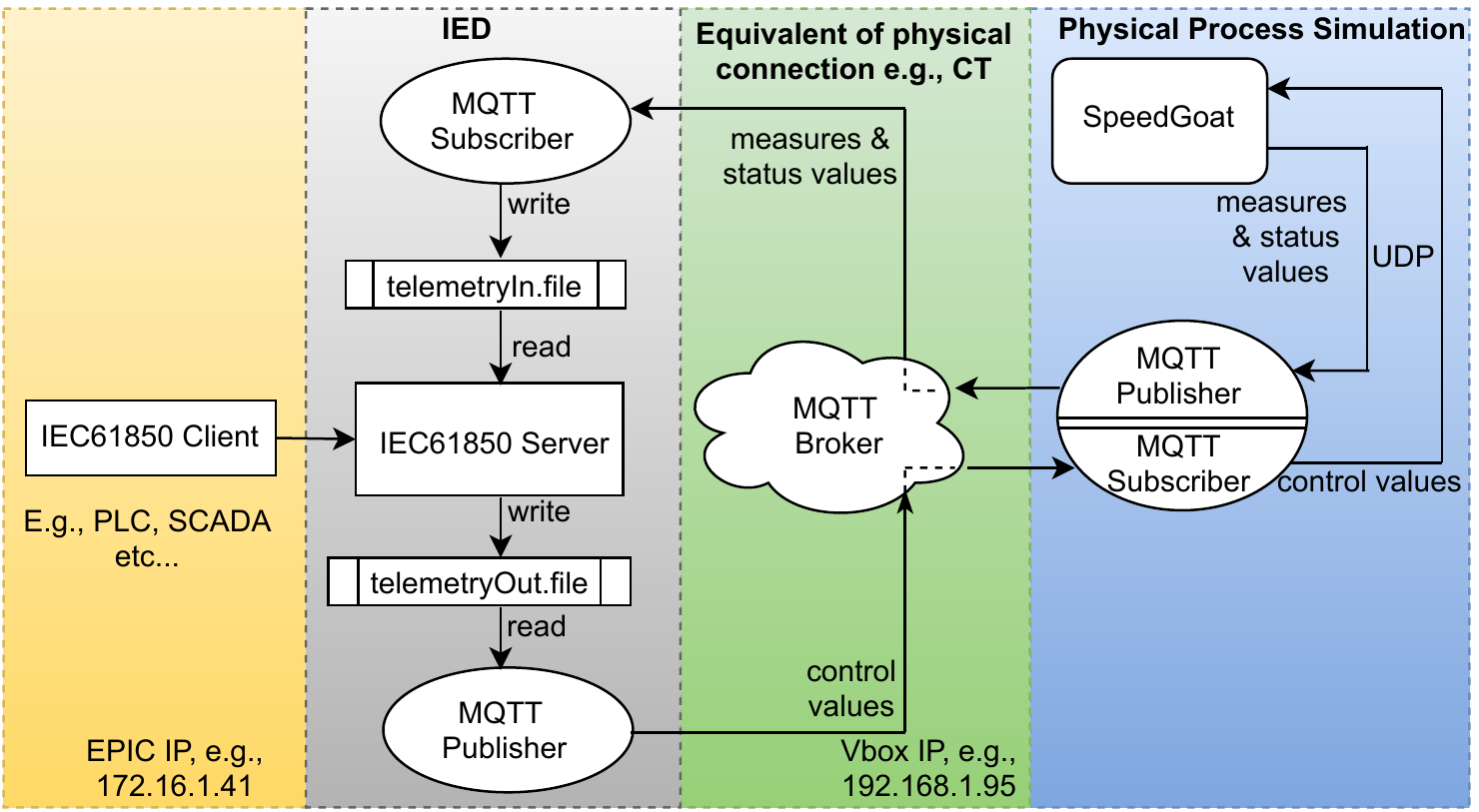}
	\caption{Illustration of process used to link the IED/AMI controllers with the physical process data on one interface and the controller network data on the other interface. \small{CT: Current Transformer, the IED rectangle shown in grey color is one of the VMs in Figure.~\ref{fig:epicsploit} say GIED1 and the two network interfaces EPIC IP and Vbox IP can be observed in IP mapping table.}}
	\label{fig:iedmodel}
\end{figure*}

\section{Architecture of the EPIC Digital Twin}
\label{sec:DTarch}

\subsection{Overview}
\label{ssec:Overview}
The architecture of the DT is shown in Figure.~\ref{DT_arch_fig}.
The physical components in Figure.\,\ref{fig:EPIC_electrical_layout} are simulated on  SPEEDGOAT and the physical signals are transferred using a data-bus to the IEDs/AMIs which act as the interface between physical and network components. The controllers required for the components in Figure.~\ref{EPIC_network_fig} are simulated using a standard LENOVO laptop\footnote{Intel i7-6600 CPU\@ 2.60GHz, 8 GiB RAM and 500 GB of SSD storage.} 
with the aid of virtual machines (VMs) running on the host machine.
Different software are used to achieve the functions of firmware, control logic and communication protocols.
We have integrated AD and Attack Launcher AL on the VMs implementing the switch/router functionalities. Different software used for achieving the functions of the physical test-bed in discussed in Section.~\ref{ssec:softcomponents} and the design consideration is discussed in Section.~\ref{ssec:design_con}.

\subsection{Software to Simulate System \& Communication Components}
\label{ssec:softcomponents}

\noindent\textbf{Physical Components and Electrical Connections}: The physical components in actual test-bed along with the electrical connections were simulated using MATLAB-SIMULINK\footnote{The model is available as a supplementary file and can executed on SIMULINK Real-time Desktop environment with relevant settings.} in real-time mode. The connection to controller simulation is achieved through an UDP data-bus. Two separate python programs handle the data flow between the simulation and rest of the system and are executed on the development computer that is connected to the SPEEDGOAT real-time machine.

\noindent\textbf{PLCs, IEDs and Network Switches}: The controllers representing PLCs, IEDs and Network switches are simulated on headless Ubuntu 18.04.5 LTS virtual Machines. We used a single operating system (OS) for scalability purposes and avoid the need for any associated license issues. However, if the users have access to the actual operating system or firmware of the respective controller it is feasible to replace the current OS with the actual  OS and firmware of the respective controllers. 
We used Virtual-Box as the hypervisor for creating and running the VMs. Our host machine also runs on Ubuntu 18.04.5 LTS.

\noindent\textbf{Run-time:} The run-time for the PLCs and IEDs are created using Node-RED platform. Node-RED is a flow-based development tool for visual programming developed originally by IBM for wiring together hardware devices, APIs and online services as part of the Internet of Things. Node-RED provides a web browser-based flow editor, which can be used to create JavaScript functions. Node-Red functions as the central point for coordinating all the servers, clients, protocol implementation and other functionalities of the respective controllers such as IEDs and PLCs (more details on NODE-RED in Section~\ref{ssec:ProgrammingFlexibility}).

\noindent\textbf{AMI Meters:} Ubuntu 32-bit docker containers are used for simulating the AMI controllers (this could also be replaced by VMs), The implementation is similar to PLCs and IEDs in other aspects.

\noindent\textbf{Network:} Open virtual switches (OVS) is used for implementing the network of the actual test-bed in twin. We use Generic Routing Encapsulation (GRE) tunnels for IP assignment and network routing.

\noindent\textbf{SCADA:} The SCADA functionalities i.e., supervisory control and all data acquisition are implemented using Node-Red. Clients for respective protocols, and the server for Open Platform Communications (OPC) \footnote{OPC is used to enable the communication between historian and HMI, similar to the actual test-bed} are co-ordinated using Node-red. The IEC clients are implemented using C and OPC server using python script.

\noindent\textbf{Historian}: Historian is implemented using Node-Red and SQLite database. The historian also provides visualization for the data similar to the actual plant.

\noindent\textbf{IEC61850}: The protocol IEC61850 used in the actual plant is implemented using \textit{\textbf{libIEC61850}} C programming library. Dedicated servers and clients are created for each controller, e.g., a PLC or an IED. As mentioned before, the co-ordination is carried out using Node-Red which also provides the run-time functions.

\noindent\textbf{OPC protocol}: The OPC protocol used for data transfer from SCADA to Historian (same as the actual plant) is implemented using python scripts and Node-Red.

\noindent\textbf{Attack Designer and Attack Launcher}:	The AD is implemented on the host machine using python scripts and it integrates the ALs implemented on individual switches to realize the attack intended. The ALs rely on Ettercap\textcolor{blue}{~\cite{ettercap}} API for implementing the packet manipulations.

\subsection{Design Considerations}
\label{ssec:design_con}

Figure.~\ref{fig:iedmodel} presents the illustration on how the physical process data is made available to the IEDs/AMIs and how other controllers on the network communicates with IEDs/AMIs. Two different network interfaces are used to achieve this implementation, i.e., Vbox IP and EPIC IP in Figure.~\ref{fig:epicsploit}. We leverage on Message Queuing Telemetry Transport (MQTT) protocol to enable the connection between the physical process and the IEDs/AMIs. MQTT is a publisher-subscriber modelled network protocol for communicating between devices~\cite{mqtt5}, and highly suitable for non-synchronous communication that is required for a data-bus.

\subsubsection{Communication Protocols}
We categorize the protocols into two groups, i) the protocols used in the actual test-bed, and ii) The protocols which are used to achieve other functions (say data-bus to represent physical connections). This section gives a brief overview of the protocols used in the DT. 

With respect to the choice of information transfer protocols, we highlight the features and properties of two widely used network communication models --- the client/server model and the publisher/subscriber model. Next, we use it as a reference to choose an approach best suited to meet the network requirements. 
Further, we look at approaches (MQTT, IEC61850 and GOOSE) under these two models and compare it to bring out their features and properties (see Table \ref{tab:comparison}).
 
 \setlength{\tabcolsep}{0.4cm}
\begin{table}[t]
	\vspace{-0.2cm}
	\centering
	\footnotesize
	
	\begin{tabular}[t]  {l c c c c c c }
		\toprule
		\textbf{\specialcell{Approach\\}} & \trot{\textbf{Client/Server}} & \trot{\textbf{\specialcell{Publisher/Subscriber}}} & \trot{\textbf{TCP/IP}}  & \trot{\textbf{Ethernet}} & \trot{\textbf{Hard Real Time}} &  \trot{\textbf{\specialcell{ Broker}}} \\
		\toprule
			
		MQTT & \hfil\xmark & \hfil\cmark & \hfil\cmark  & \hfil\xmark & \hfil\xmark &  \hfil\cmark\\
	
		IEC61850 & \hfil\cmark & \hfil\xmark & \hfil\cmark & \hfil\xmark & \hfil\xmark & \hfil\xmark\\
		
		GOOSE & \hfil\xmark & \hfil\cmark & \hfil\xmark & \hfil\cmark & \hfil\cmark & \hfil\xmark   \\

		\bottomrule
	\end{tabular}
	\vspace{5pt}
	\caption{A comparison of features and properties of network protocols used in digital twin.}
	\label{tab:comparison}
	\vspace{-0.3cm}
\end{table}

\begin{figure}[!htb]
	\centering
	\includegraphics[width=\linewidth]{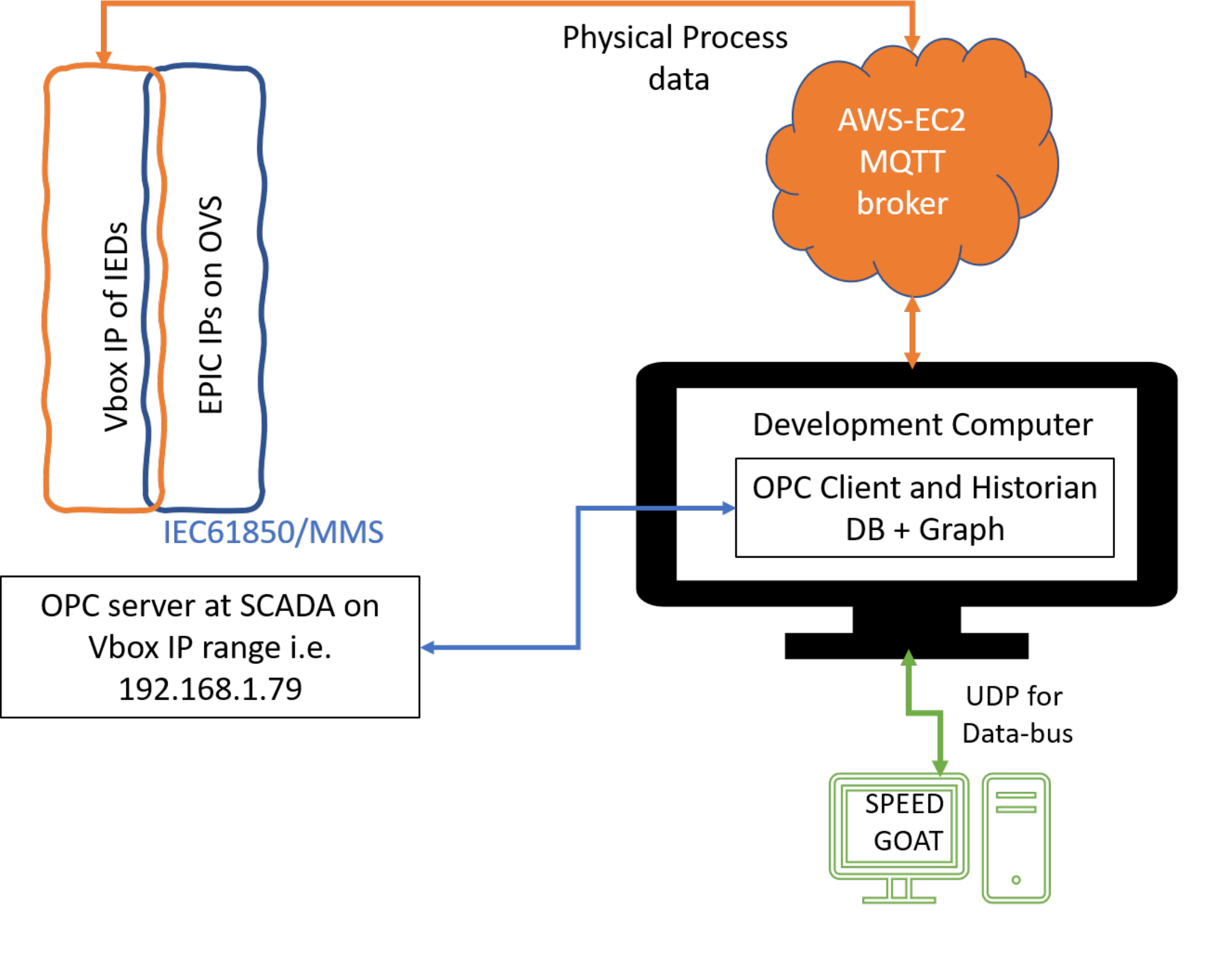}
	\caption{Connection from physical simulation on SPEEDGOAT to development computer for AWS MQTT connection and OPC data viewing on Historian. Refer Figure.~\ref{fig:epicsploit} for controller network and the details on IP mapping with vbox IP, and Figure.~\ref{fig:iedmodel} for detailed configuration of a single IED.}
	\label{fig:endtoend}
\end{figure}

\begin{itemize}
	\item \textbf{MQTT} 
	is not used in the physical test-bed, it is used in DT to achieve the data-bus functions. In MQTT, the devices are connected through an intermediate centralized broker (e.g., a host running on a cloud). A device is able to subscribe to  topic(s) on the broker. The topic is a string used to tag the communication of interest to the device.

	\item \textbf{IEC61850 manufacturing message specification (MMS)} uses Client/Server for real time communication of messages between networked devices or computers~\cite{sisco1995}. The physical test-bed uses MMS for communication between SCADA, PLCs and IEDs. MMS is used for control, measurement and status read.  
	Unlike a publisher/subscriber model that involves multiple communication endpoints, a client/server model uses a 1:1 communication at the protocol level. Our DT also uses MMS for communication.
	
	\item \textbf{Generic Object Oriented Substation Events (GOOSE)} is an IEC61850 defined control and measurement communication model in real time (typically under 4 milliseconds) over electric substation networks~\cite{Kriger2013}. In the physical test-bed GOOSE is used for communication between the IEDs. 
	GOOSE is used for time critical communication between IEDs to alert of anomalies and take immediate measures to protect the physical device(s).
	Our DT also  uses GOOSE for communication.
	
\end{itemize}

\subsubsection{Device Communication on the Network}
The main control devices in our system are typically IEDs and PLCs. \textit{Communication wrappers} are used in our DT. The IEDs on our DT  uses MMS and GOOSE similar to the physical test-bed.
The IEC client is able to directly  fetch  and set values on the server (see Figure~\ref{fig:iedmodel}).
The IEC61850 server communication is wrapped using a MQTT subscriber and publisher at the data in and out endpoints respectively. The wrapper allows for multi-cast communications between devices without disturbing existing methods used in communication.

\subsubsection{Representing Physical Connections}
As mentioned above, none of the test-beds available in literature considers the interface between the physical signals and the controllers such as IEDs/AMIs. In many cases the physicals signals are represented by IEC61850 GOOSE\footnote{Generic object oriented substation events.} signals, though some of the modern substations could be represented using such architecture, most of legacy and non-modern smart grid systems cannot be replicated. We use a dedicated data-bus (see Fig.~\ref{DT_arch_fig}) for representing the physical connections such as current/voltage sensors (transformers) to IEDs/AMIs. 

\subsubsection{Eliminating Specialized Hardware}
Though we have used SPEEDGOAT for real-time simulation, it is feasible to run the simulation on a general purpose computer with Simulink Desktop real-time. Such a setup is feasible because we use UDP packet communication for data bus (may also be deployed using serial ports). Hence, replication of the DT does not require any specialized hardware, and is significantly cheaper and scalable.

\subsubsection{MQTT vs MODBUS}
Other protocols such as MODBUS were also explored for data-bus and we observed that the performance was not affected by the choice. However, MQTT with a cloud-broker has the capability to enable multiple users to use the same simulation platform given the input/output (I/O) list is available. For example, researchers from any part of the world would be able to control the SPEEDGOAT simulation in our lab, if they have access to the I/O list and our cloud-broker's IP address.

The detailed architecture of the DT from end to end is shown\footnote{Red-dotted lines will be explained in Section.~\ref{ssec:diference}} in Figure.~\ref{fig:endtoend}. A detailed document explaining the step-by-step process to recreate the DT is also available in the supplementary files \footnote{The procedure will be updated on acceptance, as we want to protect the procedure until the paper is made public}.

\section{Case Study}
\label{sec:casestudy}
\begin{figure*}
	\centering
	\includegraphics[width=\linewidth]{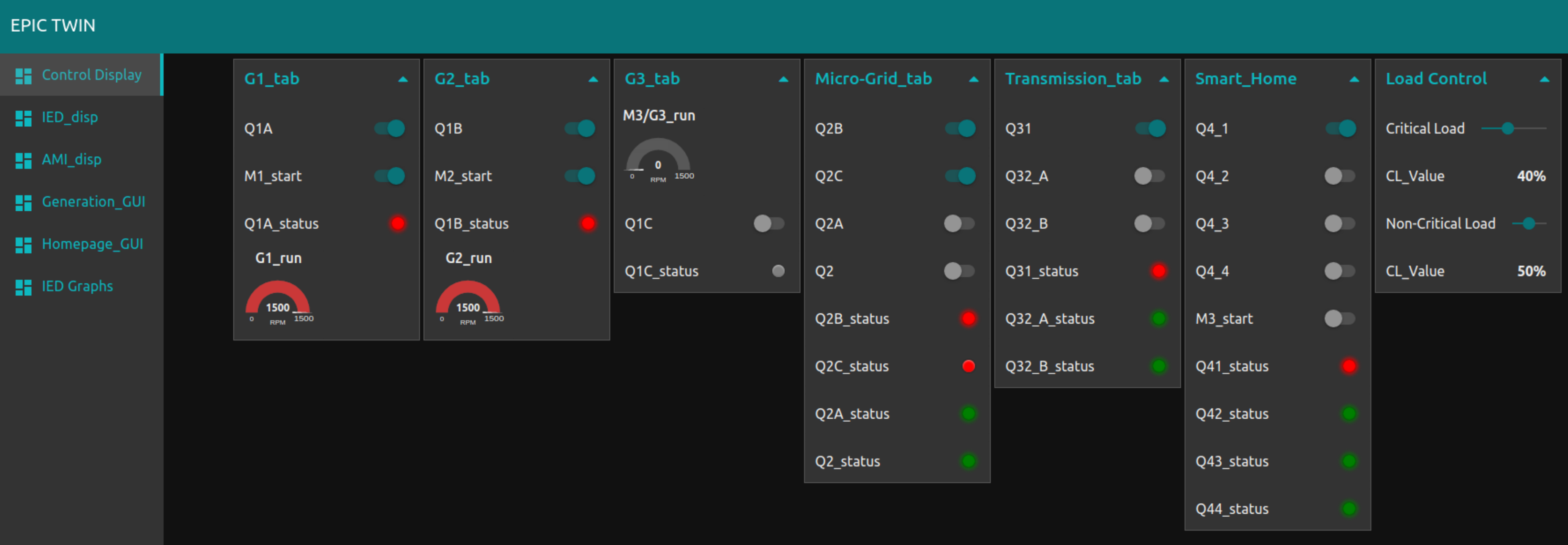}
	\caption{SCADA GUI before the attack.}
	\label{fig:controldisplay_a}
\end{figure*}

As mentioned in Section.~\ref{sec:background}, the use case of the DT is for cyber security studies. Hence, in this section, we demonstrate a complete and coherent attack on the smart home zone. Consider a load (e.g., a aggregated heat pump) attached to the smart home zone consisting of multiple homes. The operator is able to  read the SCADA  display GUI (see Fig.~\ref{fig:controldisplay_a}) by logging in via an  application. The status indicator will show the position of the switch controlling the aggregated heat pump. The color red indicates the status OFF and green represents ON. The owner (victim) sends a control signal to turn off the pump.

However, a man-in-the-middle (MITM) attacker mounts an attack to  maintain the status of the heat pump as ON.
To cover their trail,  a command to display the status of the pump as OFF is passed to the SCADA. As a result, the  operator sees the status of the pump as OFF. 
This may be insufficient to go unnoticed when the current and voltage values are also part of the owners GUI visuals. It would appear that a higher current is still being drawn by the load despite appearing to be turned off. 
To correct it, the attacker also spoofs the measurement values to replicate the conditions when the load was disconnected.

\begin{figure}[!htb]
	\centering
	\includegraphics[width=\linewidth]{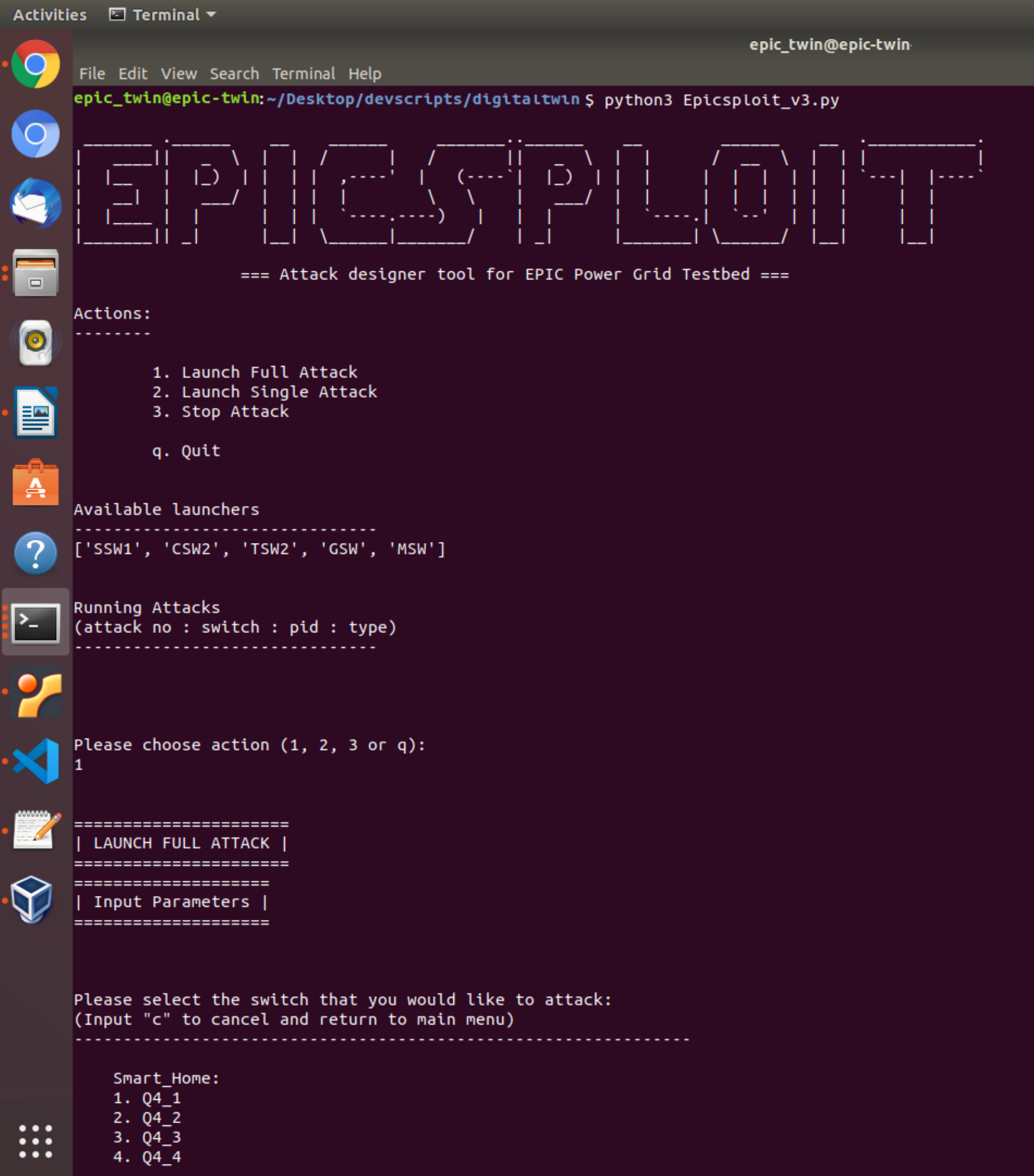}
	\caption{An attacker using the EpicSploit tool to spoof control, status and measurement values send to the SCADA.}
	\label{fig:controldisplay_b}
\end{figure}

To demonstrate this, we developed EpicSploit (see Fig.~\ref{fig:controldisplay_b}) --- an attack designer tool for EPIC's~\cite{Sridhar2019} DT. The goal of the tool is to exhibit fast and easy active man-in-the-middle attacks on specified points within EPIC twin via a command line interface that users can conveniently navigate and spoof values to represent an attacker. EpicSploit is a generic and composable  tool not limited to the attack scenario described above. As seen in Fig. \ref{fig:epicsploit}, EpicSploit is a plug-and-play tool running on the EPIC's DT. The  tool is made up of two main components: the attack designer and the attack launcher. 
The AD (for this implementation) is located on the SCADA workstation, and will provide a simple user interface on which users can design and run attacks on EPIC DT as an attacker. The AL may also run on any other component between the attacked points. The AL can be on the controllers, say PLC to realize attacks originating from  edge devices, arising from its vulnerabilities.

Once the attacker enters their commands to EpicSploit, it makes API calls to the attack launchers located on the network switches within EPIC DT. Here, virtual machines represent the network switches. These attack launchers run the  scripts  to carry out active  MITM attacks on the target points. Using EpicSploit--- measurement, control and status values passing through the network are manipulated.  For the smart home attack described earlier, the EpicSploit tool allows an attacker to select a ON/OFF switch. E.g., Q4\_1 from Smart Home (See Fig. \ref{fig:controldisplay_a}). Further, the attacker selects the payload (whether it stays on or is turned off). The tool will run an attack command on the target network switch and reflect on the SCADA, the readings corresponding to the attack payload. 

When the operator has sent a command to that switch, EpicSploit will automatically run status and measurement attacks to spoof the respective values associated with the target switch. The operator is made to believe a successful response for the command sent had the desired effect. However, the tool had intercepted and modified the command in the background. The users can develop and test their defense (say an intrusion detection system (IDS)) mechanism using EpicSploit. The IDS could either be based on design centric approach (using OPC or MMS data), data centric approach (using OPC or MMS data), process based, network based or a combination of multiple methods. The readers are directed to refer~\cite{Sridhar2019,adepu2019attacks,kandasamy2019investigation} for examples. The IDS can be deployed on any of the VMs in the DT. However, IDS implementation was not considered in the scope of this paper. \textbf{\textit{A screen video recording of the attack is available in one of the supplementary files.}}

\section{Discussion}
\label{sec:discussion}
\subsection{Programming Flexibility}
\label{ssec:ProgrammingFlexibility}

\textbf{Node-Red} is a visual programming tool~\cite{nodered} used for wiring together (but not limited to) hardware devices. It provides online services (e.g., MQTT-in and MQTT-out) and supports a wide variety of APIs for applications that are either pre-built with the installation or downloadable from its library. The applications  are also highly configurable from the GUI.
Node-Red is ideal for building and editing small to medium sized workflow. It empowers users by supplying a powerful substitute to command-line batching through process workflow visualization. The availability of a large library of ready to use nodes and its flexibility to insert user built nodes, and to interconnect them makes it a powerful tool for real time workflow automation, monitoring and debugging.
Since headless VMs are used in our implementation, we have installed Node-Red (with web server support) on the VMs. A web browser (on the host machine) connects to the Node-Red server on the VM and acts a visual interface to control and display the execution of the binaries located on the VM.
Our implementation of the DT makes extensive use of Node-Red for workflow automation. 

\subsection{Messaging Scalability}
\textbf{MQTT} is found  to develop communication latencies when a large number of devices are subscribed to the broker or when there are multiple publishers (or a combination of both).
From prior experimentation in Aichernig and Schumi~\cite{Bernhard2018}, it was  observed that MQTT implementations such as Mosquitto (used in this work) have a reasonably low latency of 100 milliseconds with around 90\% of the 100 mosquitto devices able to meet this latency threshold (for the usage profiles tested). A latency of 100 to 1000 milliseconds is considered acceptable for natural progression of the task (e.g., time to loading webpages or changing views)~\cite{railmodel}. The EPIC DT uses a total of 20 VMs, with most of them running a MQTT publisher/subscriber. No visible latencies were observed from the use of MQTT in our DT.

\subsection{Virtual Machine vs Container}
\label{ssec:vmvscontainer}
We attempted to  implement the entire system using containers in place of VMs to further limit resource consumption. However, due to unavailability of containers that can be used for MITM attacks and due to filtering limitation of Ettercap for docker container data traffic, container based implementation was found to be not practical for DT focusing on cyber security studies. However, if the DT is used for other applications say `Energy Management', container based implementation may be a prudent solution. 

\subsection{Differences between the DT and physical test-bed}
\label{ssec:diference}
Though in an ideal scenario the expectation is to create a exact equivalent of the physical test-bed, there are conditions that limit certain features to be reproduced. The differences along with practical constraints that resulted in the deviation are listed below.
\begin{compactenum}[i)]
    \item \textbf{HSR} ring is used in the physical test-bed for creating the internal (within the dotted box) and external rings as shown Figure.~\ref{EPIC_network_fig}. HSR rings provide protection against a single point of failure in the data network. However, in our DT --- since we rely on OVS to recreate the network, recreating HSR is not feasible at this point of time. Hence, we have not removed the links shown in red-dotted line in  Figure.~\ref{fig:epicsploit} to enable the stability of the virtual network created.
    \item Each zone in the physical network has \textbf{two switches} (e.g., Micro-grid has MSW1 and MSW2) in the the physical test-bed. In the DT, we have implemented the functions of these two switches on a single VM to optimize the computation resources. 
    \item \textbf{AMI meter data} in the physical test-bed is made available to SCADA in the following sequence, AMI meter (DLMS\footnote{(Device Language Message Specification)} protocol) 	$\rightarrow$ RaspberryPi ( DLMS to  MODBUS protocol) 	$\rightarrow$ AMI Switch 	$\rightarrow$ PLC (MODBUS - IEC61850 MMS) 	$\rightarrow$ SCADA. In our DT, the data is directly made available in IEC61850 MMS format from the docker containers to SCADA.
    \item \textbf{The passive loads} in physical test-bed are controllable through discrete values such as 5kW, 2kW, 1kW (also equivalent kVAr). In our DT, it is controllable from 0 to 100\% in steps of 10\%.
    \item \textbf{The water test-beds} are fixed loads in the DT whereas they are real dynamic loads in the physical test-bed.
    \item \textbf{Solar PV and Battery system} is modeled as PQ source in the DT without any power electronic components. This is to optimize the computation load on the real-time machine.
    \item \textbf{I/O list} of the DT is similar to the physical test-bed in the functions but actual I/O list is not used due to confidentiality constraints. The I/O list corresponds to the actual variable names used to control and monitor the plant.
\end{compactenum}

\section{Related work}
\label{sec:related}
Security and privacy issues in smart grids have been discussed in literature~\cite{Yan2012,Liu2014,Wang2013,Komninos2014}.
We discuss the existing test-beds whose focus is on smartgrid security.
Finally, we compare them with the EPIC digital twin to highlight its benefits.
 
\textbf{Cybersecurity test-bed comparisons:}
A number of  hardware~\cite{NREL,jeju,IITGALVIN}, hybrid~\cite{Multiphysics} and simulator~\cite{Stanovich2013,Tran2013,Ingram2011} CPS smart grid test-beds are available.
However, the test-bed platforms of interest are those where their targeted research area is cybersecurity.
The PRIME test-bed~\cite{prime} uses a similar architecture to our twin, the differences are three fold i.e., no attack tool (similar to our AD and AL) is available, needs specialised hardware (Compact Rio) and software (such as Labview) for implementation and it is not a digital twin of any existing system. The DETER test-bed~\cite{Mirkovic2010} and National SCADA test-bed Program ~\cite{INL2009} are hardware cybersecurity test-beds. Due to their physical nature, it is not straightforward to explore different architectures, modify or scale. The University College Dublin test-bed~\cite{Hong2011} is a hybrid where the power system is simulated  whereas the IEDs and circuit breakers at the substation level are physical devices. Their power system simulator and user interface communicates via Object Linking and Embedded Process Control (OPC). 
IEC61850 based protocols are used between substation user interface and IEDs. The use of physical devices limits its scalability and adds to the cost of the test-bed. The TASSCS test-bed~\cite{Mallouhi2011}  uses a number of commercial tools to build their simulator. They use OPNET tool to simulate computer networks and PowerWorld simulation system to simulate operation segments in the electric grid, and Modbus RSim to simulate Modbus servers.  

The use of commercial tools as opposed to open source restricts its audience. 
 The work in Yang et al \cite{Yang2015}, focuses on fuzz testing for IEC61850 based IEDs but use physical IEDs and network switches.
Texas A\&M University supplies ~\cite{Chen2014} a real-time  cyber-physical security test-bed. It uses real time digital simulator (RTDS) to simulate the power system and its controllers, OPNET to simulate network communications and Labview PXI plaltform to bridge RTDS and OPNET in real time. The proposed architecture shows the use of physical IEDs and physical networking devices. This results in scalability and cost issues associated with physical devices.
The work in Dayal et al~\cite{vscada} presents VSCADA, a virtual SCADA infrastructure. However, they use OPC for communication both at the SCADA front and  backend.
SCADASim~\cite{Queiroz2011}, a real time test-bed simulator has 3 main components ---  a message synchronization scheduler, a communication mechanism, and simulation objects to represent external components in the simulation environment. It is built in top of OMNET++ (a discrete event simulator). However, MODBUS/TCP simulated protocols cannot be used as its socket stack is not supported and is replaced by INET, a library that simulates the TCP/IP protocol.
 The work in Koutsandria~\cite{Koutsandria15} focuses on network intrusion detection and uses a rasberry pi as a network tap --- typically used to monitor traffic between two systems connected via a physical cable.
 
 In the case of EPIC DT, all network switches, components and their interconnections are virtual. I.e., the network switches, IEDs and PLCs run on VMs. 
 Th EPIC DT requires no additional physical resources to carry out MITM attacks or to include an intrusion detection system.
 The use of libiec61850 C library ensures that the protocol stack remains consistent with the real test-bed.

\section{Conclusions}
\label{sec:conclusions}

 In this paper a DT for an existing ICS plant was presented and in the developed DT the ICS components such as IEDs, PLCs and network switches are realised on virtual machines, and AMIs on docker containers. The interconnecting virtual network was established using OVS. As a result, the software components are open source, easy to scale and the trace from a packet sniffer is similar to that on a physical test-bed. In comparison, other network simulators may not trivially support the required communication stack and the sniffer trace may not be identical to a physical test-bed. This causes overheads in security testing and training. Unlike other simulated test-beds reviewed, the EPIC digital twin  was designed to closely resemble the outcome of security testing carried out on a physical test-bed. The developed test-bed was also designed with AD and AL capable of  enabling systematic testing and bench-marking of IDS that are developed using data-centric and design centric approaches. EPICTWIN in conjunction with the existing digital test-beds,  would be a useful tool for generating process and network data that might be impossible with real plants or actual test-beds.
 Overall, our simulation required two standard off-the-shelf laptops and is an attractive low cost solution for the grid security development and testing, and can be extended to other types of ICS.

\section{Acknowledgment}
\label{sec:Acknowledgment}
This work was supported in part by the National Research Foundation (NRF), Prime Minister’s Office, Singapore, under its National Cybersecurity R\&D Programme (Award No. NRF2018 - NCR - NSOE005 - 0001) and administered by the National Cybersecurity R\&D Directorate.

\bibliographystyle{elsarticle-num}
\bibliography{paper}

\end{document}